\documentclass[printer]{aa}  
\usepackage{graphicx}
\usepackage{txfonts}
\usepackage{textcomp}
\usepackage{hyperref}
\hypersetup{
  linkcolor  = blue,
  citecolor  = blue,
  urlcolor   = blue,
  colorlinks = true,
}

\begin{document}

   \title{The need for new techniques to identify the high-frequency MHD waves of an oscillating coronal loop}

   \author{Farhad Allian
          \and
          Rekha Jain
          }

   \institute{School of Mathematics and Statistics, The University of Sheffield, S3 7RH, UK \\ 
   \email{fallian1@sheffield.ac.uk}}

   \date{Received X; accepted X}

  \abstract
   {Magnetic arcades in the solar atmosphere, or coronal loops, are common structures known to host magnetohydrodynamic (MHD) waves and oscillations. Of particular interest are the observed properties of transverse loop oscillations, such as their frequency and mode of oscillation, which have received significant attention in recent years because of their seismological capability. Previous studies have relied on standard data analysis techniques, such as a fast Fourier transform (FFT) and wavelet transform (WT), to correctly extract periodicities and identify the MHD modes. However, the ways in which these methods can lead to artefacts requires careful investigation.} 
   {We aim to assess whether these two common spectral analysis techniques in coronal seismology can successfully identify high-frequency waves from an oscillating coronal loop.}
   {We examine extreme ultraviolet (EUV) images of a coronal loop observed by the Atmospheric Imaging Assembly (AIA) in the 171 {\AA} waveband on board the Solar Dynamics Observatory (SDO). We perform a spectral analysis of the loop waveform and compare our observation with a basic simulation. }
  {The spectral FFT and WT power of the observed loop waveform is found to reveal a significant signal with frequency ${\sim}2.67$ mHz superposed onto the dominant mode of oscillation of the loop (${\sim}1.33$ mHz), that is, the second harmonic of the loop. The simulated data show that the second harmonic is completely artificial even though both of these methods identify this mode as a real signal. {This artificial harmonic, and several higher modes, are shown to arise owing to the periodic but non-uniform brightness of the loop.} We further illustrate that the reconstruction of the ${\sim}2.67$ mHz component, particularly in the presence of noise, yields a false perception of oscillatory behaviour that does not otherwise exist. {We suggest that additional techniques, such as a forward model of a 3D coronal arcade, are necessary to verify such high-frequency waves.} }
  {Our findings have significant implications for coronal seismology, as we highlight the dangers of attempting to identify high-frequency MHD wave modes using these standard data analysis techniques. }


	\titlerunning{New techniques required to identify the high-frequency MHD waves of an oscillating coronal loop}
	
\maketitle

\section{Introduction}

Coronal loops are often observed to vacillate in the plane of the sky. In particular, transverse coronal loop oscillations are one of the most widely studied phenomena in the solar atmosphere. To date, substantial efforts involving theory, simulations, and observations have been made to extract the inaccessible yet imperative physical properties of the loops by magnetohydrodynamic (MHD) seismology. The commonly accepted model describes the transverse perturbations as the free, instantaneous, and non-axisymmetric (kink) eigenmodes of a cylindrical waveguide \citep[e.g.][]{Roberts1984}. However, alternative frameworks consisting of an entire magnetic arcade also exist \citep{HindmanJain2014, Hindman2015,Hindman2018}.  A recent review of MHD waves and oscillations can be found in \cite{Nakariakov2020}.

Initially, the spatially resolved motion of coronal loops was discovered by the Transition Region and Coronal Explorer (TRACE) in extreme ultraviolet (EUV) wavelengths as periodic displacements induced by flaring activity \citep[][]{Aschwanden1999}.  Since then, further studies have shown that these periodicities range from a few to several tens of minutes \citep{Aschwanden2002} and are found to be strongly correlated with the length of the loop \citep{Goddard2016}. However, their excitation mechanism is debated. \cite{Hudson2004}  suggested that a fast-mode shock wave expelled from a flaring epicentre may be important. In a more recent observational catalogue of 58 events, \cite{Zimovets2015} found that the majority (57 events) of transverse oscillations are perturbed from an equilibrium by nearby impulsive eruptions, instead of shock waves. \cite{Zimovets2015} also reported that 53 of these events were associated with flares, which may imply a relationship between these two types of magnetic activities.   {Although it is agreed that flares play an important role in the excitation of loop oscillations}, why certain periodicities are enhanced over others remains unknown.

In addition to impulsive loop oscillations, observational efforts have revealed ambient, small-amplitude oscillations that persist in the absence of any (obvious) driver without a significant decay \citep[][]{Wang2012, Nistico2013,Anfinogentov2015, Duckenfield2018}. Their periods of oscillation are similar to the impulsive regime and different segments of the loop, from the footpoints to the apex, have been shown to oscillate in phase \citep{Anfinogentov2013}. Even though it is agreed that their source of oscillation must be small scale and likely broad band \citep{HindmanJain2014}, their precise excitation mechanism is also unknown and several theories have been proposed.  \cite{Nakariakov2016} considered small-amplitude oscillations as self-oscillatory processes due to the interaction between quasi-steady flows at the loop footpoints. \cite{Afanyasev2020} suggested that a broadband frequency-dependent driver at the loop footpoints can lead to the excitation of several waveguide modes. A recent and comprehensive 3D MHD simulation by \cite{Kohutova2021} demonstrated that a harmonic footpoint driver is not a prerequisite for the excitation of loop oscillations. 

Both regimes of transverse loop oscillations are found to predominantly exhibit the fundamental mode, that is, with nodes at the footpoints of the loop and an anti-node at its apex. Higher-order modes can also be excited; however, their presence is rare in comparison \citep[e.g.][]{Verwichte2004}. The primary interest in understanding the mode of oscillation of a loop lies in its seismological capability when used in tandem with 1D models \citep[][]{Andries2005}. Often, the approach is to compute the ratio of a loop's observed fundamental period $P_1$ to its nth overtone, that is $P_1/nP_n$ \citep{Duckenfield2018,Duckenfield2019}.   {For a dispersion-less oscillation, it is believed that any deviation of this period ratio from unity may suggest a plasma density stratification within the loop \citep{Andries2009}. On the contrary, \cite{Jain2012} demonstrated using sensitivity kernels for a cylindrical waveguide that this period ratio only contains broad spatial averages of the wave speed and is highly insensitive to the loop density. \cite{Jain2012} suggested that it is necessary to obtain inversions of several frequency modes and additional non-seismic observations are needed to infer about the density.}

A clear verification of the existence of higher-order modes however can be difficult to detect from observations. It is well understood that the emission from coronal plasmas in EUV wavelengths is optically thin. Observationally, this means that a particular coronal loop is preferentially illuminated owing to a superposition of intensities along the line of sight from local heating processes. This unfortunate property of coronal wavelengths introduces a particularly challenging task of disentangling the emission from nearby loops with fidelity, including for mode verification \citep[][]{DeMoortel2012}. To date, observational studies of transverse loop oscillations have excluded the high frequencies in favour of the signal whose intensity is brightest in the image foreground for seismological purposes with 1D models \citep[e.g.][]{Nakariakov2001}.  {Two common techniques for separating a dominant periodic signal from its background include filtering frequencies within the data spectrum \citep[e.g.][]{Terradas2004, Morton2012} and tracking the peak intensity as a function of time \citep[][]{Li2017}}.   {The foreground signal is calculated by creating a time series of the maximum intensity of a loop (typically modelled as a Gaussian) within a time-distance map and is often interpreted as the resonant kink modes of a 1D isolated waveguide \citep[e.g.][]{Pascoe2016, Pascoe2020}. 
        
However, there are now multiple lines of evidence that show oscillations are not only confined to the visible loop and therefore a propagation of waves across the magnetic arcade must be present \citep[e.g.][]{Jain2015, Allian2019, Conde2020}.  Noting the cross-field propagation from observations, \cite{HindmanJain2014} and \cite{Hindman2015} argued that the true nature of a coronal loop wave cavity is multidimensional, and an examination of the power spectrum of the waveform is imperative for understanding the origin of signals from observational data. In particular, \cite{HindmanJain2014} demonstrated that the presence of ambient, high-frequency signals from a coronal arcade may be indicative of a stochastic excitation mechanism. Within their 2D framework, a source consisting of a broad range of frequencies embedded in the background can excite fast MHD waves, which are trapped standing waves longitudinal to the field, while propagating perpendicular to the arcade. Moreover, \cite{HindmanJain2014} postulated that an observed impulsive waveform is therefore a superposition of waveguide modes from an ambient background source and an impulsive driver.  Thus, in principle, the power spectrum of an observed coronal loop time series contains seismic information about the excitation mechanism and the arcade waveguide.

In a broader context, several studies have also revealed that the power spectrum of dynamical processes in the solar corona can follow a power-law distribution \citep{Aschwanden2011, Auchere2014, Ireland2015,Kolotkov2016}. \cite{Aschwanden2011} initially proposed that the power-law behaviour of random processes in solar time series could be due to the superposition of several small energy deposition events. \cite{Auchere2014} analysed 917 events of EUV intensity pulsations and found power laws with frequencies ranging from $0.01 - 1$ mHz. Similarly, \cite{Ireland2015} revealed power-law properties from active regions in the 171 {\AA} and 193 {\AA}  wavebands. These authors argued that the power-law spectra of coronal time series must be considered for the automation of detection algorithms, the correct interpretation of the underlying physical processes and coronal seismological inferences. \cite{Ireland2015} also cast doubt on how frequencies can be extracted from data due to the unknown efficacy of a priori defined background noise models.  While much attention has been given to the power-law distribution of solar time series and their associated noise models, a careful investigation into how the standard analysis techniques may lead to artefacts in the context of coronal seismology remains to be satisfactorily addressed.

The hypothesis of our study is as follows: If a coronal loop is observed to oscillate with a single frequency, then diminishing the strength of that signal should accentuate the presence of ambient wave frequencies, if they exist. We test this by performing a spectral analysis on the waveform of an observed coronal loop oscillating with a single frequency and compare our results with a synthetic loop embedded in a background of noise. We show that the identification of wave frequencies from an observed oscillating loop is non-trivial and the shape of the waveform indirectly influences the detected frequencies. In Section \ref{Data} we provide an observational overview and describe our data analysis techniques. In Section \ref{Results} we present our results. Finally, we discuss the implications of our findings and summarise key results in Section \ref{Discussion}.

\section{Data and analysis methods}
 \label{Data}

\subsection{Observational overview}

\begin{figure}
        \includegraphics[width=8cm]{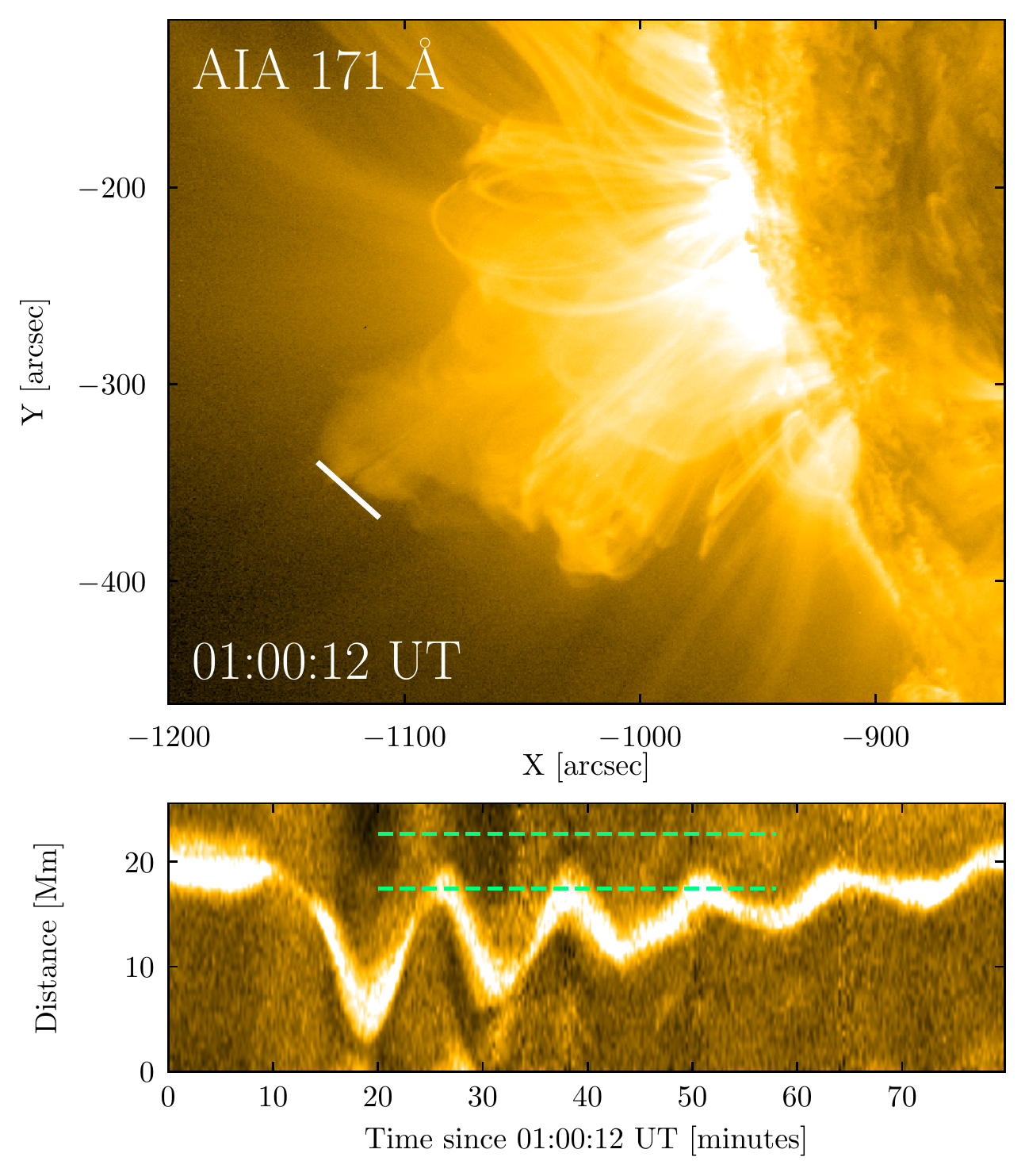}
        \caption{Top panel: Snapshot of our region of interest observed by SDO/AIA at 01:00:12 UT. The origin of our slit corresponds to $(x,y) = (-1111, -366)$ arcsec. Bottom panel: Time-distance map of a bright coronal loop exhibiting clear transverse oscillations with a dominant periodicity of ${\sim}13$ min. {The dashed green lines highlight the two pixel locations used for further analysis.} }       
        \label{Figure1} 
\end{figure}

For our study, we utilise the same observational event as in \cite{Allian2019}. In their work, the authors developed a novel spatio-temporal analysis method to examine the periodicities of faint and bright transverse loop oscillations. \cite{Allian2019} found periodicities ranging from 9-13 min between loops in proximity, which was {later} confirmed by \cite{Pascoe2020}.  {\cite{Pascoe2020} also reported the absence of a higher frequency component from these loops. In this work, we use a combination of spectral techniques to investigate whether such high-frequency oscillations exist.}

The coronal loop of interest, also referred to as Slit 1 in \cite{Allian2019}, was observed on 2014 January 27 off the southeastern limb of the Sun with the Atmospheric Imaging Assembly \citep[AIA;][]{Lemen2012} on board the Solar Dynamics Observatory \citep[SDO;][]{Pesnell2012}. Our analysis utilises AIA 171 {\AA} data, with a spatial resolution of $0.6''$ pixel$^{-1}$ (${\approx}0.435$ Mm) and temporal cadence of 12 s. This dataset was chosen due to the high-quality and well-contrasted observable conditions of a bright coronal loop in the image foreground. A nearby M1.0 class flare initiated around 01:05:12 UT and perturbed the apex of several loops. The top panel of Figure \ref{Figure1} shows the region of interest in which transverse oscillations took place at the arcade apex and the slit (white line) used to create the time-distance map. The resultant time-distance map we use for our study is shown in the bottom panel of Figure \ref{Figure1}. Prior to the flare onset, the loop appears as a compact shape with non-uniform brightness, which spans a projected distance of approximately $3$ Mm within the slit. Thereafter, the loop exhibits a clear transverse oscillation with a dominant periodicity of ${\sim}13$ min.  

\subsection{Spectral methods}

Our aim is to investigate the frequency content of the raw data in search of a high-frequency component superposed onto the bright loop shown in the bottom panel of Figure \ref{Figure1}. To do this, we firstly employed a fast Fourier transform (FFT) of the time-distance map. The data is apodized in time and space with a $\cos^2$ curve to mitigate frequency leakage before computing the FFT along each dimension. Other window functions were tested for verification and produced similar results; which, for brevity, is omitted. The power spectrum is then calculated as the magnitude squared of the FFT spectrum and is normalised with respect to the signal variance, ${\sigma_0}^2$ \citep{Torrence1998}.  It is worth noting that, apart from the standard procedure outlined above, no detrending or smoothing is applied to the data.  Time-distance maps of loop oscillations are often averaged within neighbouring pixels to increase the signal-to-noise ratio of the data  \citep[e.g.][]{Pascoe2016}, and therefore any high-frequency signal is decimated, causing bias in the interpretation of the power spectrum. In this work, we do not perform any smoothing and the full cadence and pixel resolution of the original data is retained to obtain any fine-detail intensity variation of waves that may be present. This data has temporal ($\nu_{max}$) and spatial Nyquist ($k_{max}$)  frequencies of approximately 41.67 mHz and 1.15 Mm$^{-1}$, respectively.

To assess the significance of wave components in the FFT spectrum, we carefully selected an appropriate background noise model. The theoretical study of \cite{HindmanJain2014} demonstrated that a white noise source can excite fast MHD waves, which travel throughout the arcade waveguide. As the aim of our study is to isolate such background frequencies, which are likely to have a constant amplitude for all frequencies, we use a null hypothesis test based on a theoretical white noise spectrum and calculate the $5 \%$ significance threshold ($95 \%$ confidence level) from the data \citep{Torrence1998}. Frequencies with power greater than the significance threshold are identified as real signals from waves traversing the coronal loop. Further justification for choosing such a background noise model is corroborated with a basic simulation, which is described in Section \ref{Synthetic}. 

Furthermore, we validate our initial FFT analysis and account for any non-stationary signals that may be present within the data by performing a wavelet analysis \citep{Torrence1998}. This technique is often preferred over a traditional FFT analysis in coronal seismic studies owing to its ability to distinguish both the frequency and temporal content of a given signal \citep[e.g.][]{Duckenfield2019, Pascoe2020}. 
We also relax the assumption of a white noise background in our wavelet analysis to account for any frequency dependence from the data and estimate the corresponding $95\%$ confidence levels of the wavelet power.

\section{Results} \label{Results}

\subsection{Observed waveform} \label{Observed}

\begin{figure}[t!]
        \includegraphics[width=8cm]{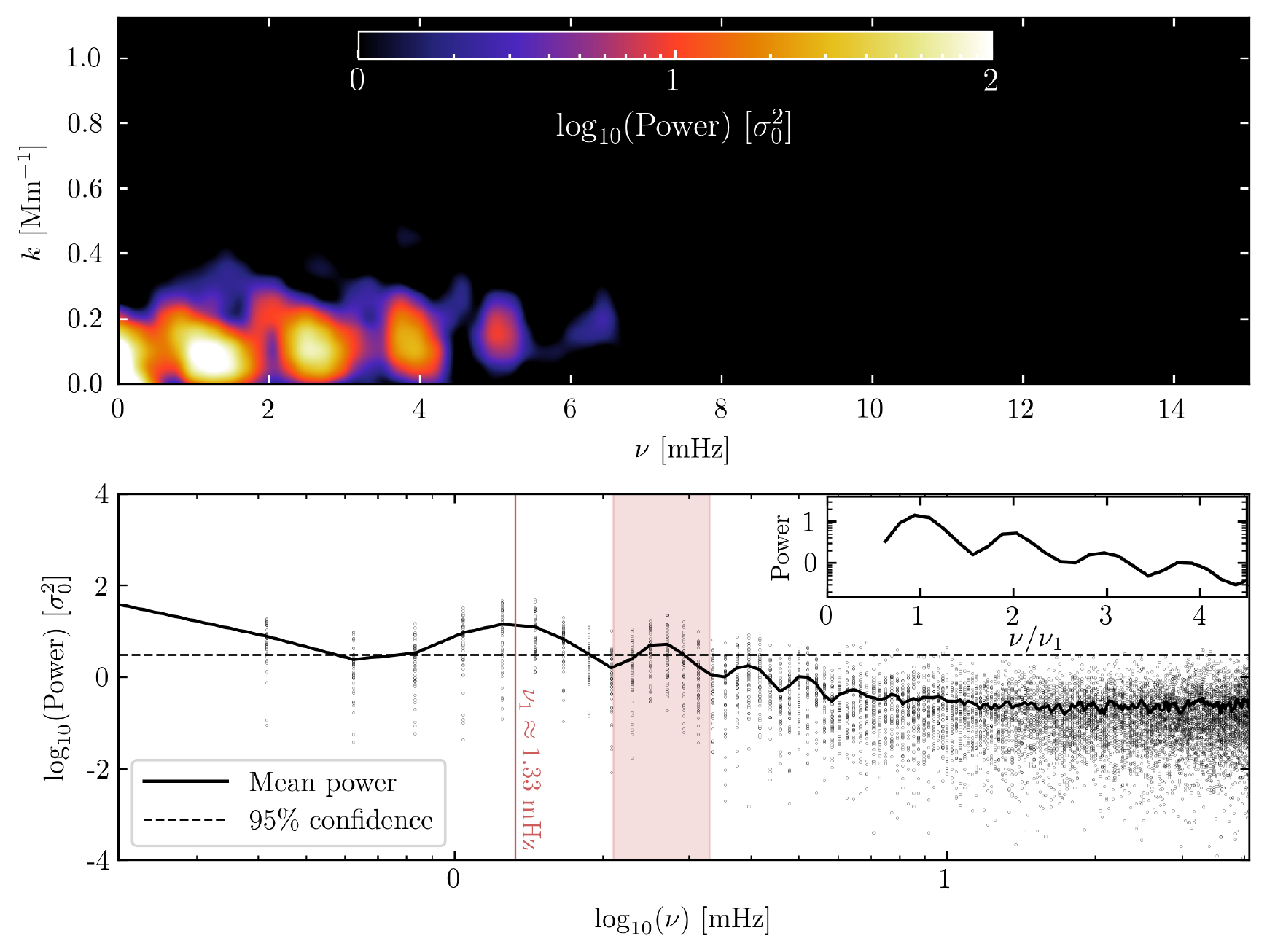}
        \caption{Top panel: Two-dimensional FFT power spectrum of the time-distance map shown in the bottom panel of Figure \ref{Figure1} as a function of spatial frequency, $k$, and temporal frequency, $\nu$.  The power is normalised with respect to the variance of the time series signal within each pixel. The peak power occurs at a fundamental frequency of $\nu_1 \approx 1.33 $ mHz, corresponding to the dominant 13 min period of the loop.  Signatures of high-frequency components are also present at $\nu > \nu_1$ with decreasing power. Bottom panel: Log-log plot of the FFT power as a function of frequency. The black dots represent the distribution of spectral power within each pixel domain. The solid black line indicates the mean power and the dashed line indicates the $95\%$ confidence level estimated from the theoretical white-noise spectrum. {The shaded red region highlights a secondary significant frequency band with a peak at ${\sim} 2.67$ mHz. } }      
        \label{Figure2} 
\end{figure}

\begin{figure*}[t!]
        \centering
        \includegraphics[width=18cm]{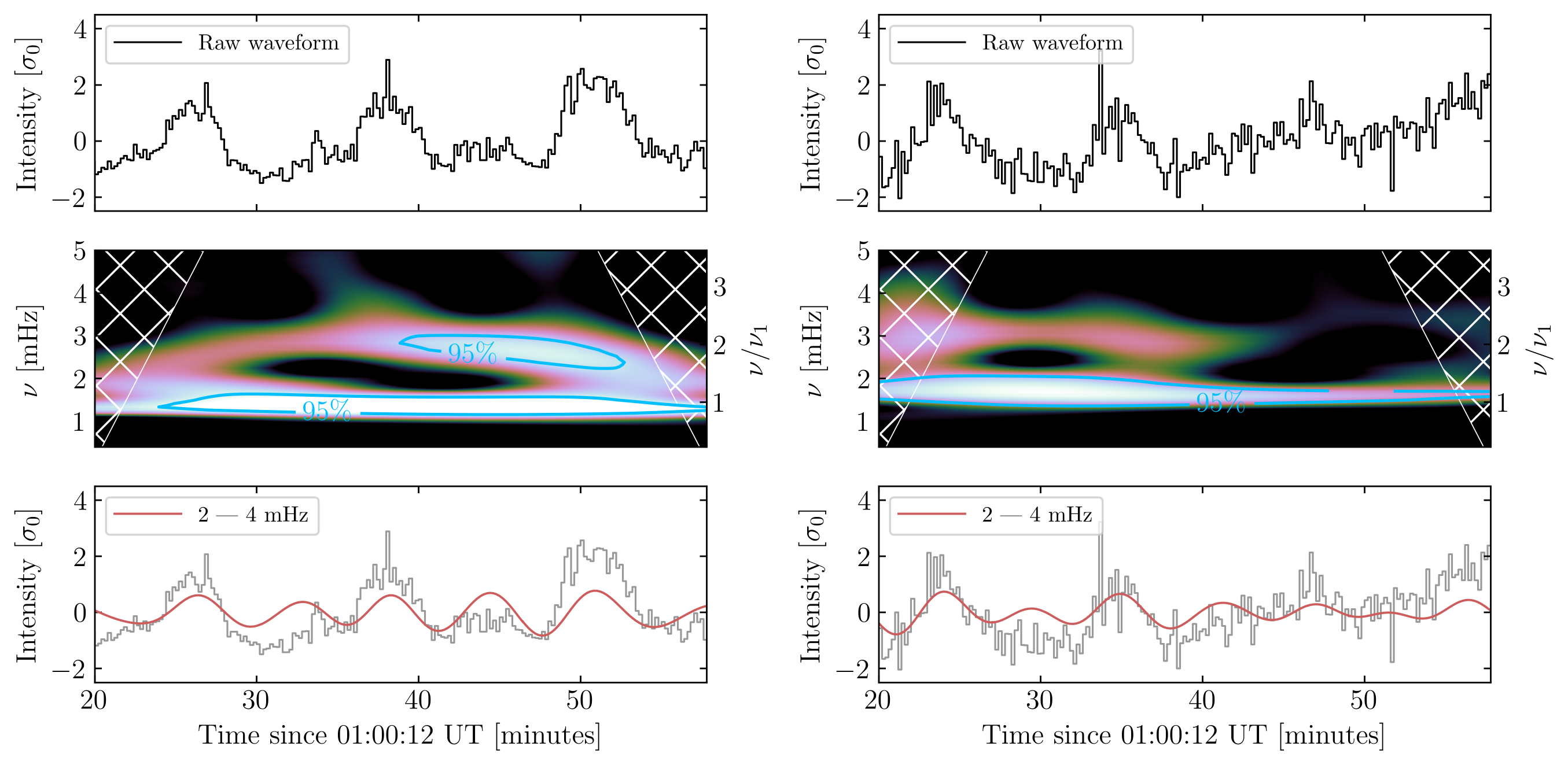}
        \caption{{Analysis of the observed waveforms from the bright loop (left) and the faint oscillatory region (right) highlighted by the green lines in the bottom panel of Figure \ref{Figure1}}. The top panels show the raw waveforms normalised with respect to the standard deviation. Ambient signatures of high-frequency signals are present in addition to the dominant $13$ min oscillations. The middle panels shows the wavelet power of the raw signals. For the bright loop, two statistically significant oscillatory frequencies are present at approximately 1.33 mHz and 2.67 mHz, which is highlighted by the blue contours. The bottom panels show the 2-4 mHz bandpass filtered signals (red) overlaid onto the observed waveforms (grey). There are signatures of ${\sim}6$ min periodic signals superposed on the raw waveforms.}       
        \label{Figure3} 
\end{figure*}

The top panel of Figure \ref{Figure2} shows the resultant 2D FFT power spectrum of the oscillating coronal loop data as a function of temporal frequency, $\nu$, and spatial frequency, $k$. As expected, the peak power occurs at the fundamental temporal frequency $\nu_1  \approx 1.33$ mHz, corresponding to that of the dominant period of the loop ($\sim$13 min).  {Surprisingly, we also find power at frequencies greater than the fundamental mode of the loop following a near-linear ridge, suggesting that the observed $13$ min waveform is almost dispersion-less. 
The relatively low power of the flare-induced high frequencies ($\nu \gtrsim 10$ mHz) is currently unclear and has been previously reported \citep[][]{Liu2011}. One obvious possibility is that the signals produced by the flaring driver at these frequencies exist, but are much weaker in strength.}  {\cite{HindmanJain2014} demonstrated that the presence of ambient power at high frequencies can be attributed to fast MHD waves propagating across finer spatial structures of a coronal arcade, and could be due to ambient stochastic sources embedded within the background.}

The log-log FFT power of each pixel domain and their mean is shown in the bottom panel of Figure \ref{Figure2}. The distribution of power in each pixel appears flat at high frequencies.  Noticeably, the peak frequencies are equidistant with decreasing strength, as we would expect to observe from a signal whose harmonics have been excited.  This pattern is seen more clearly in the inset plot of Figure \ref{Figure2}, showing the mean power distribution as a function of a frequency ratio (normalised with respect to $\nu_1$).  Evidently, the power peaks at several integer multiples of the fundamental frequency.} \cite{Pascoe2020} state that they found no evidence of a higher frequency component in search for overtones from these loops. However, we can clearly see that signatures of such high-frequency harmonics exist within the raw data. A white noise hypothesis test supports our initial claim that the second harmonic (${\sim}2.67$ mHz) is significant and present within the waveform, based on the estimated $95\%$ confidence level. At this initial stage of analysis, it could be possible to identify this frequency as the higher-order overtone of the loop.  A more thorough treatment, of course, requires a multitude of time-distance maps spanning the projected loop in search of a clear phase difference. From this method, we were also unable to find convincing evidence of phase change from this loop.  As shown in this work, however, the intensity variations corresponding to the harmonics of an oscillating loop cannot be easily distinguished within coronal time series. 

To study the observed waveform shown in the bottom panel of Figure \ref{Figure1} in greater detail, we considered two cases: a pixel located on the bright loop where the transverse oscillations appeared most prominent, and another located away from the loop where the transverse oscillations were not as visible. Our objective is to isolate any high-frequency waves that may be superposed on the bright loop. We now describe the first case. Following the oscillation start time at approximately 01:10:12 UT, the bright loop exhibits a contraction that causes it to drift off the slit before beginning its dominant $13$ min cycle. As a result, we extracted a shorter time series ($x = 18$ Mm) of ${\sim}40$ min duration starting from 01:20:12 UT, where the loop appeared most stable for three cycles before decaying and examined the raw AIA waveform. This is also motivated by our expectation of the relatively short-lived duration of high-frequency waves, whose presence can go unnoticed within longer duration time series. The top left panel of Figure \ref{Figure3} shows the extracted waveform, which is normalised with respect to the standard deviation ($\sigma_0$) of the signal. In addition to the pronounced 13 min waveform, there is a clear presence of ambient, high-frequency jitter embedded within three cycles of the flare-induced waves \citep[see][]{Allian2019}. The wavelet power of the observed waveform is shown in the middle of Figure \ref{Figure3} as a function of frequency and time. We find that the wavelet transform (WT) of this time series also results in two statistically significant frequencies at ${\sim}1.33$ mHz and its second harmonic (${\sim}2.67$ mHz), in accord with our initial FFT analysis. At present, we posit that the ${\sim}2.67$ mHz frequency component with relatively low power is either due to the presence of an ambient stochastic driver superposed onto the dominant flare-induced waves or a weak signature of the overtone of a neighbouring loop. In an attempt to isolate the ${\sim}2.67$ mHz component, we applied a bandpass filter to the raw AIA waveform between 2-4 mHz. The bottom left panel of Figure \ref{Figure3} presents the filtered signal that we suspect to be embedded within the loop, which is over plotted to provide a comparative visualisation relative to the total observed waveform. Clearly, these ${\sim}2.67$ mHz (6 min) oscillations permeate throughout the coronal loop, however, their presence within the raw data is practically indiscernible as a result of the high-power contribution from the dominant ${\sim}1.33$ mHz component. A similar waveform is shown in the right panel of Figure \ref{Figure3} for the faint region case ($x = 23$ Mm) in the same duration. Signatures of ${\sim}2.67$ mHz frequencies are still present in the wavelet power; however, this component falls below the $95\%$ confidence interval likely because of the relatively poor signal-to-noise ratio of the waveform. The frequency content for all pixels can also be inferred by assessing the distribution of FFT power shown in the bottom panel of Figure \ref{Figure2}.
        
The amplitude of the reconstructed signal from the bright loop is approximately $25 \%$ of the total intensity. We estimated the uncertainty of this reconstructed signal by calculating the expected noise level of our observed waveform. Following \cite{Yuan2012}, the noise level in the 171 {\AA} waveband is calculated according to Poisson statistics to be $(0.06I + 2.3)^{1/2}$, where $I$ is the overall intensity in units of DN. For typical intensity values within the bright loop (${\sim}150$ DN), this yields an error of ${\pm}3$  DN. This value is around $\pm 0.1 \sigma_0$ in the standardized units of intensity. The intensity of the reconstructed signal is around two to three times higher than this estimated noise level, and we are left to believe that the 2.67 mHz component represents some real mechanism. However, as we demonstrate with a basic simulation in the following section, this component is an artefact that arises because of the non-uniform brightness of the observed coronal loop.

\subsection{Synthetic waveform} \label{Synthetic}

\begin{figure}
        \includegraphics[width=8cm]{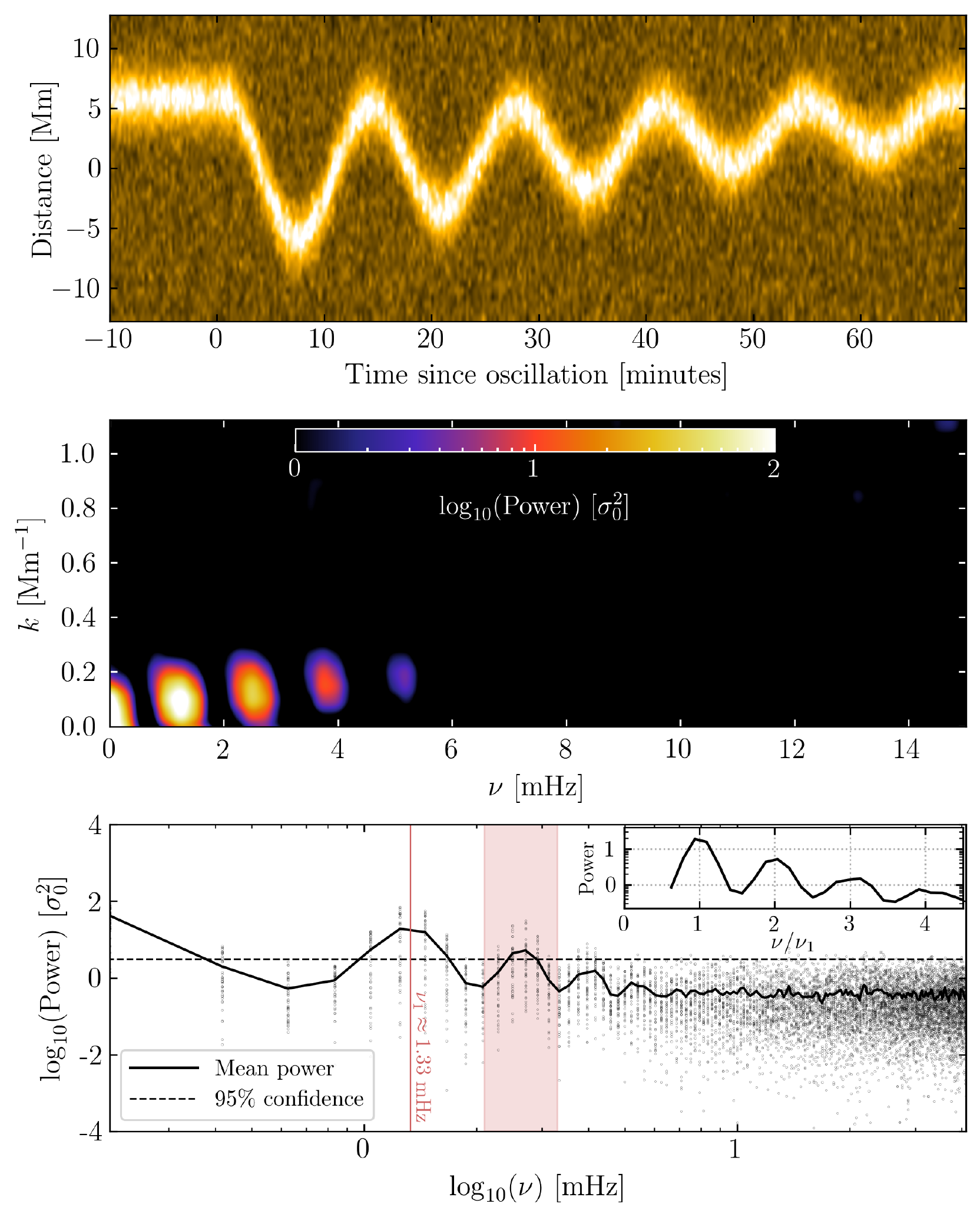}
        \caption{Basic simulation of a coronal loop that suddenly oscillates with a single periodicity of 13 min, embedded within a uniform background of noise. Top panel: Synthetic time-distance map of a loop oscillating with a single periodicity that exhibits a slow exponential decay and linear drift along the slit. Poisson noise and white noise sources have been included. Middle panel: Two-dimensional FFT power of the simulated waveform shown in the above panel. The pattern of frequencies is akin to the observed waveform shown in Figure \ref{Figure2}, that is the power peaks at $1.33$ mHz and at several of its harmonics. Bottom panel: Log-log plot of the FFT power of the synthetic waveform. The symbols and colours are the same as in Figure \ref{Figure2}.}       
        \label{Figure4} 
\end{figure}

\begin{figure}
        \includegraphics[width=8cm]{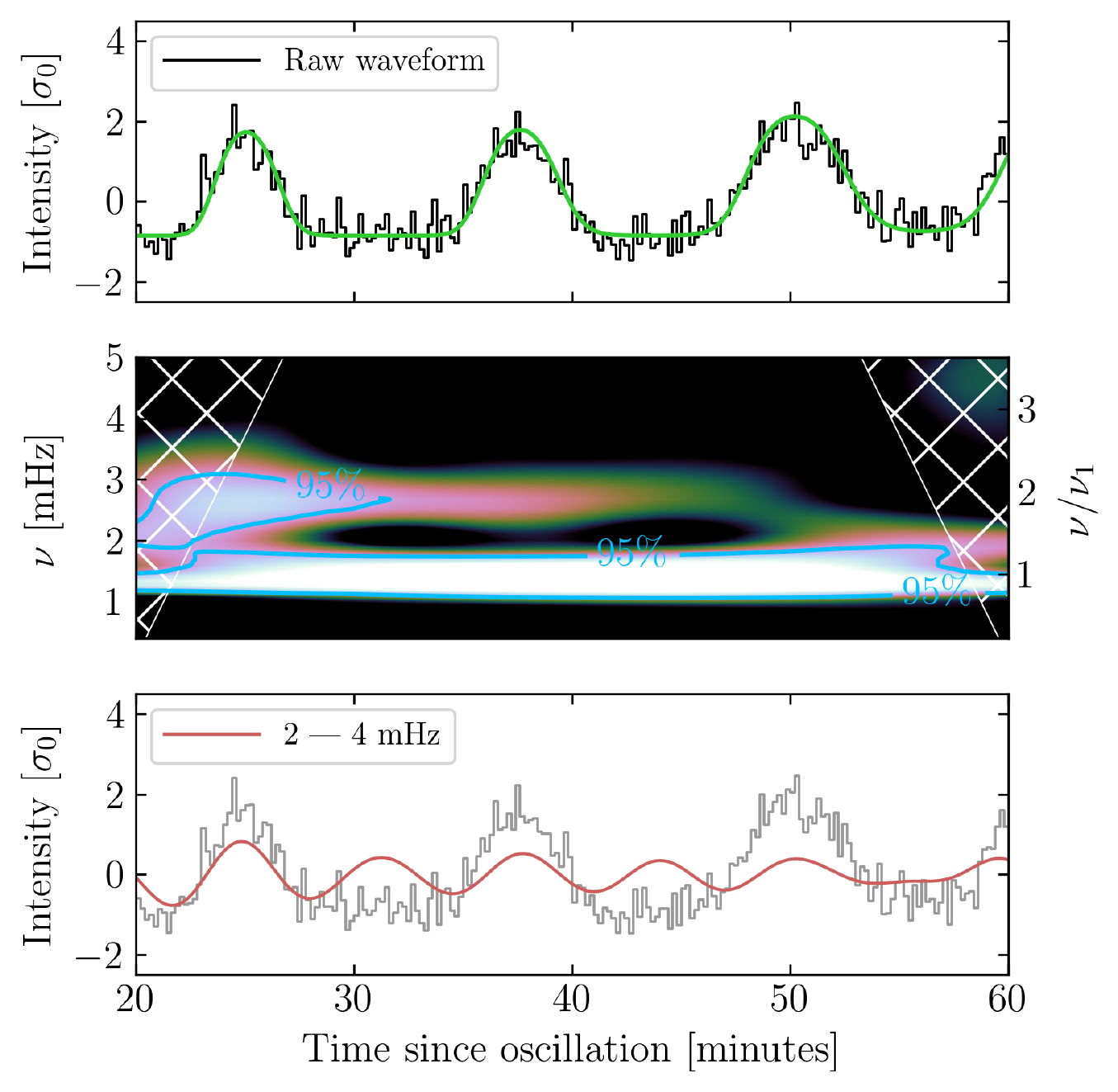}
        \caption{Same as Figure \ref{Figure3} but for the synthetic loop waveform. The green line overlaid in the top panel represents the 13 min synthetic waveform in the absence of any noise. The presence of the second harmonic (${\sim}2.67$ mHz component) in the wavelet power is artificial and arises as a result of the periodic but non-uniform (Gaussian) brightness of the loop.}       
        \label{Figure5} 
\end{figure}

Now that we have carefully examined the spectral content of the observed waveform, we perform a basic simulation of a coronal loop oscillating with a single periodicity. Our goal is to better understand the presence of the harmonics observed in the raw AIA data. To this end, we synthesised a time-distance map of a coronal loop oscillating with a single periodicity (13 min) observed at the AIA cadence and pixel resolution. The loop is simulated in a $45 \times 400$ domain, which represents a 25 Mm long slit observed for a duration of 80 min, respectively. For simplicity, we assume that the loop cross-sectional brightness is Gaussian across the slit and that the amplitude and width of the loop are constant throughout its lifetime. The loop oscillates sinusoidally with a single frequency as

\begin{equation}
y(t) = \xi_0 \cos(2\pi \nu t + \phi)e^{-t/\tau }+ \textnormal{trend},
\end{equation}
where $\xi_0$ is the displacement amplitude of the loop, $\nu$ is the frequency of oscillation, and $\phi$ is the phase offset. The sinusoid is modulated with an exponential decay term with a decay time $\tau$, as is commonly expected for flare-induced transverse waves. A linear trend is also included across the slit to account for the observed growth in the time-distance map in Figure \ref{Figure1}, perhaps resulting from a change in the magnetic field topology from the flare. Finally, we include a constant background intensity that consists of contributions from both Poisson noise and artificial white noise at their expected levels. The former is added to mimic photon noise within each AIA pixel and the latter to account for any additional broadband source that may be present \citep{HindmanJain2014}.

Figure \ref{Figure4} demonstrates the hierarchical process we used to infer about the waveform of the synthetic loop.  The top panel shows the time-distance map of a loop oscillating with a single periodicity embedded within a uniform background of noise. The displacement amplitude of the loop is 5 Mm with a 13 min periodicity that suddenly oscillates. The loop displacement decays with an e-folding time of $45$ min. The middle panel shows the 2D FFT power spectrum of the synthesised time-distance map. We immediately find that the synthetic power spectrum reflects a striking resemblance to that of the observed loop in Figure \ref{Figure2}. Similarly, the bottom panel of Figure \ref{Figure4} shows the log-log plot of the FFT power as a function of temporal frequency, where there is a significant power enhancement around $2.67$ mHz in addition to the expected $1.33$ mHz. The high-frequency tail of the power spectrum is also uniformly distributed, reinforcing our claim that the observed waveform is likely to be dominated by a white-noise source. A close inspection of the bottom panel in Figure \ref{Figure4} reveals that the FFT power decreases as $\nu^{-1.5}$ for at least three harmonics before tending to a more uniform distribution at higher frequencies, suggesting a power-law model might be appropriate within the low-frequency range. This is particularly suspect since the synthetic waveform was created with only a dominant white-noise source contribution. Hence, the overall shape of a 1D FFT spectrum can be dictated indirectly by the observed shape of the waveform itself. Nevertheless, it is clear that the ${\sim}2.67$ component of the synthetic waveform is also significant and requires further investigation.

We now proceed to analyse the time series of the synthetic waveform of the bright loop in the same manner as described in Section \ref{Observed}. The top panel of Figure \ref{Figure5} shows the raw synthetic waveform in which both noise sources are included. The green lines overlaid onto the raw waveform represents that of the noise-free loop. The wavelet power in the middle panel of Figure \ref{Figure5} shows a significant power enhancement at ${\sim}2.67$ mHz throughout the duration of the time series, and the corresponding bandpass filtered signal in the bottom panel also shows signatures of ${\sim}2.67$ mHz signals embedded within the loop waveform. We reiterate that no wave frequency other than the 1.33 mHz component was included in our set-up. 

The presence of the high-frequency harmonics in Figure \ref{Figure4} and Figure \ref{Figure5} can be explained as follows. Consider the waveform within a single pixel, such as that in the top left panel of Figure \ref{Figure5}. Since the loop has a non-uniform width that is defined by its brightness, or density inhomogeneity, then the time series from a single pixel contains information about the loop periodicity and the lifetime of brightness within that pixel. In other words, we may think of the resultant waveform as convolution with a Dirac comb spaced every ${\sim}13$ min in time and a Gaussian shape that is defined by the loop of width $\sigma_x$ traversing a single pixel. Therefore, the power spectrum of such a waveform yields another Dirac comb with a frequency of $\nu_1 \approx 1.33$ mHz multiplied by another Gaussian of width proportional to $\sigma_x^{-1}$ and the spectral power of the time series decreases like a Gaussian (further discussed in Appendix \ref{A1}). As a result, frequencies greater than $\nu_1$ are essentially aliases of this mode. In reality, an observed coronal waveform contains more time-dependent features that are non-trivial to simulate and may cause further difficulty in interpreting the frequency content, for instance, from a change in width or periodicity of the structure. The wavelet power of the waveform contains only up to the second harmonic, as opposed to the FFT spectrum, which is primarily due to the resolution of the wavelet filter itself \citep[see][]{Torrence1998}.

\subsection{Comparison of the two waveforms}

\begin{figure}
        \includegraphics[width=8cm]{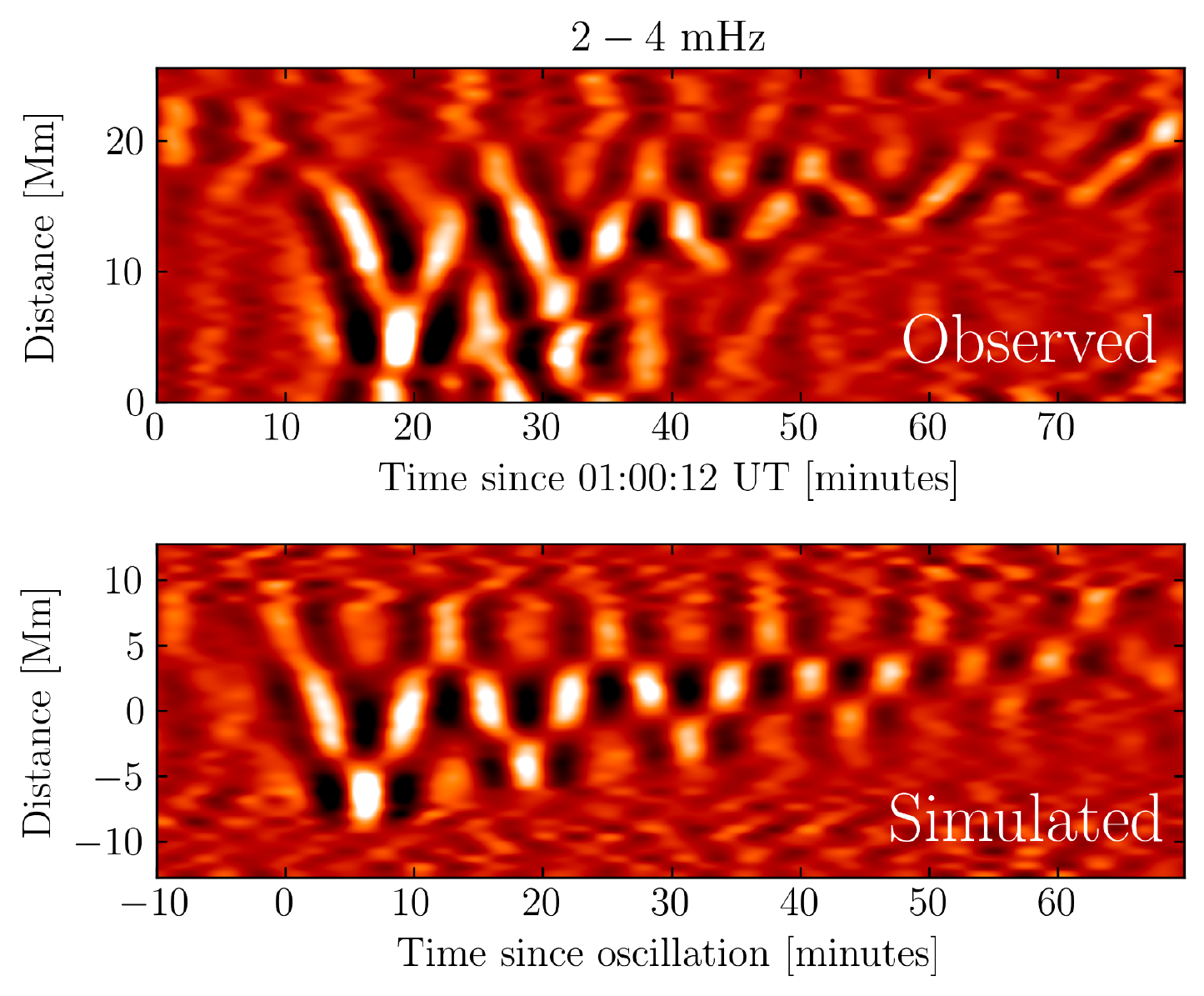}
        \caption{Reconstructed coronal loop waveforms (2-4 mHz) of the observed (top) and synthetic data (bottom). The dominant oscillatory (1.33 mHz) pattern is still visible despite diminishing its frequency. Signatures of small-scale oscillatory behaviour are also present. In both cases, these oscillations are completely artificial and arise owing to the non-uniform brightness of the loop in the presence of noise.   }       
        \label{Figure6} 
\end{figure}

It is now instructive to return to the observed and synthetic waveforms to better visualise the high-frequency oscillatory behaviour (or lack thereof) in the entire spatial domain. Figure \ref{Figure6} shows a comparison of the $95\%$ significant 2-4 mHz signals from the observed (top) and synthetic (bottom) data. Visibly, the dominant 13 min component of the loop is still present in both cases despite diminishing its strength. This reinforces our previous claim that frequencies greater than the dominant mode ($\nu > \nu_1$) of the loop are essentially aliases and arise as a result of the periodic but non-sinusoidal shape of the coronal loop waveform. This result has severe consequences for coronal seismology, as we demonstrated with unambiguous examples that one can only confidently infer about the dominant frequency of a given signal. Moreover, attempts of isolating or filtering a specified frequency band of a coronal waveform can lead to artificially enhanced oscillations and cannot be used as evidence of high-frequency behaviour. The presence
of high-frequency oscillations can be supported by examining the 2D power spectrum as a function of  $k$ and $\nu$. This is because a high-frequency component of a coronal loop results in a shift in spatial frequency that is not evinced within a 1D spectrum (see Appendix \ref{A2}). To re-emphasise, our results suggest that these frequencies are artefacts that arise because of the inability of the spectral techniques to distinguish periodic but non-sinusoidal signals.

\section{Discussion and conclusions} \label{Discussion}

We have performed a spectral analysis of an observed coronal loop oscillating with a dominant periodicity of 13 min (${\sim}1.33$ mHz) in search of a high-frequency component. Using a combination of FFT and wavelet analysis, we `found' evidence of a significant second harmonic component (${\sim}2.67$ mHz) embedded within the dominant mode of oscillation of the loop. A basic simulation of the loop revealed that this component, including frequencies greater than the fundamental component, is artificial. These high-frequency harmonics arise as a result of periodic but non-sinusoidal oscillations, the shape of which is defined by the non-uniform brightness of the coronal loop itself. We argued that, with just these techniques alone, high-frequency signals from an oscillating loop cannot be identified with fidelity. We demonstrated that the power spectrum of an observed coronal waveform and its frequency dependency may be better understood by inspecting the 2D spectrum as a function of spatial and temporal frequency.  In reference to our initial hypothesis, it is now clear that diminishing the strength of the dominant frequency cannot confidently reveal the existence of high-frequency waves. Despite using two common and independent analysis techniques in coronal seismology, and adjusting our background noise assumption, these signals are still identified as real signals from both methods. In the following subsections, we explore the plausibility of detecting high-frequency oscillations and the implications for coronal seismology.

\subsection{High frequencies from a coronal loop: Real or artificial} \label{Background}

Our observed results presented in Section \ref{Observed} demonstrates ambient signatures of the second harmonic (${\sim} 2.67$ mHz) of a coronal loop superposed onto the dominant mode of oscillation of the loop, using two independent analysis techniques in coronal seismology. The FFT and wavelet spectrum identified this component as a `real' signal from their corresponding $95\%$ confidence levels. Only by examining the 2D FFT spectrum together with a basic simulation (see Section \ref{Synthetic}) were we able to rule out the possibility of a genuine signal being present. However, interestingly, the periodicity of this signal ({${\sim}6$ min) and its persistence is comparable to that of the loop several hours before the impulsive flare, which was shown to exhibit small-amplitude 9 min oscillations and may indicate a stochastic driver \citep[see][]{Allian2019}. Our interpretation is similar to that of \cite{Nistico2013} who derived an empirical model of transverse loop oscillations as the response to two distinct drivers: a continuous non-resonant source and an impulsive driver. A more rigorous mathematical framework developed by \cite{HindmanJain2014} suggested that fast MHD waves emanating from a stochastic (white) source can excite the resonant modes of an arcade waveguide \citep[see Figure 8 of][]{HindmanJain2014}. Within the 2D model of \cite{HindmanJain2014}, the primary role of the flare is to enhance the power of frequencies that are already present within the background of the arcade. Although the FFT and wavelet power of this component from the observed loop waveform (left panel of Figure \ref{Figure3}) is slightly more enhanced than that of the synthetic waveform (Figure \ref{Figure5}), we believe insufficient evidence remains to suggest the presence of a continuous, resonant source superposed on the bright loop within the duration of the flaring activity.

It is also natural to question whether these artificial harmonics arise owing to the basis functions (complex sinusoids) of the FFT and, by extension, the WT. It is well known that the FFT suffers from the distortion of non-sinusoidal signals. We speculate that a more suitable approach of analysing solar coronal waveforms could be accomplished using the adaptive (and basis independent) empirical mode decomposition (EMD) algorithm, which only a handful of studies previously have explored \citep[e.g.][]{Huang1998, Terradas2004, Morton2012, Kolotkov2016}. For instance, \cite{Terradas2004} employed EMD filtering on TRACE observations of coronal loops oscillations to obtain the spatial distribution of propagating and standing waves of periods 5 and 10 min, respectively. \cite{Terradas2004} suggested that the intensity fluctuation of the EMD filtered 10 min period may be indicative of the radial overtone of the loop being excited. While we demonstrate in this work that filtering a frequency band of an observed waveform can result in artificial frequencies, even with appropriate significance testing, it is clear that further work is needed to explore the applicability of the EMD algorithm in solar applications.  This could be the subject of future work.

\subsection{Implications for coronal seismology}

The discussion of Section \ref{Background} explores whether the background oscillatory signals of an observed waveform are genuine or not. To date, seismological inferences have heavily relied upon the standard FFT and WT techniques to extract the frequencies of oscillating coronal loops. Our results highlight the dangers of over-interpreting signatures of high frequencies from observed coronal waveforms and add to the complexity of their nature reported in previous works. \cite{Ireland2015} highlighted the importance of incorporating the power-law distribution of coronal waveforms including appropriate background noise models for the correct seismological inference and the automation for oscillatory detection algorithms. While our work has focussed on large-amplitude oscillatory signals, \cite{Ireland2015} analysed quiet-sun regions, including loop footpoints. However, several studies have shown the prevalence of small-amplitude oscillations that persist from the loop footpoints to their apex in quiet-sun regions \citep{Anfinogentov2013, Anfinogentov2015}. Recent high-resolution observations have also revealed fine-scale coronal loop strands from the Hi-C instrument, which are almost unresolvable by SDO/AIA \citep[see][]{Aschwanden2017, Williams2020b, Williams2020a} and, as a result, the emission from nearby strands are likely to contribute to the overall emission of what we perceive as the visible coronal loop. From our results presented in this work, it is now clear that the power spectrum of signals from coronal structures may consist of artefacts, such as higher harmonics and power-law behaviour solely from the width of the observed loop waveform. Although the power spectrum of small-amplitude oscillations is almost comparable to the background noise, we believe they can contribute to the overall spectrum and can result in biases, particularly when applying image processing techniques or any non-linear manipulation of the raw data. However, as we have demonstrated, it is possible to rule these frequencies out by consulting the 2D FFT and wavelet power of the waveform in tandem with a basic simulated model. We also note that while our work has focussed on the oscillation of a coronal loop, our inference applies to any transversely oscillating structure that consists of a non-uniform brightness. 

{While our basic simulation has successfully elucidated the presence of high-frequency components from the observed data, a more realistic set-up is required to model how the emission of EUV plasmas evolve in space and time. We suggest that a forward modelling approach of an entire 3D coronal arcade may be prudent to account for more complex configurations \citep[e.g.][]{Hardi2006}. Similarities from the observed and 3D modelled waveforms may then be revealed, for instance, using cross-correlation analysis. A significant improvement in the spatial and temporal resolutions of the detector may also be necessary to convincingly identify high-frequency modes from observations.}

Finally, we comment on the validity of searching for loop overtones from observations \citep[e.g][]{Pascoe2016, Duckenfield2019}. Such studies extract the dominant (foreground) signal before conducting a spectral analysis using either FFT or WTs by estimating the position of peak brightness as a function of time where the projected loop exhibits a clear phase difference \citep[also see][]{Verwichte2009}. From this, we find non-negligible signatures of up to the artificial second harmonic of the loop due to the sampling from its non-uniform brightness in the FFT power. On the other hand, the wavelet power retains the fundamental mode but generally smooths out the presence of the artificial second harmonic. Thus, from this approach, a wavelet analysis with appropriate significance testing can be suitably used in seismic studies as others have envisioned.

\begin{acknowledgements}
We thank the anonymous referee for their valuable comments that helped improve our manuscript. We acknowledge the support of STFC (UK). F.A is grateful for the STFC studentship, and would also like to thank Bryony C. Moody and Samuel J. Skirvin for insightful discussions. The data is provided courtesy of NASA SDO/AIA science teams. This research has made use of SunPy, an open-source software package for solar data analysis \citep{Sunpy2020}.
\end{acknowledgements}

\bibliographystyle{aa.bst} 
\bibliography{References} 

 \begin{appendix}

\section{Two-dimensional power spectrum a synthetic coronal  loop oscillating with a single frequency} \label{A1}

 \begin{figure*}[t!]
        \centering
        \includegraphics[width=18cm]{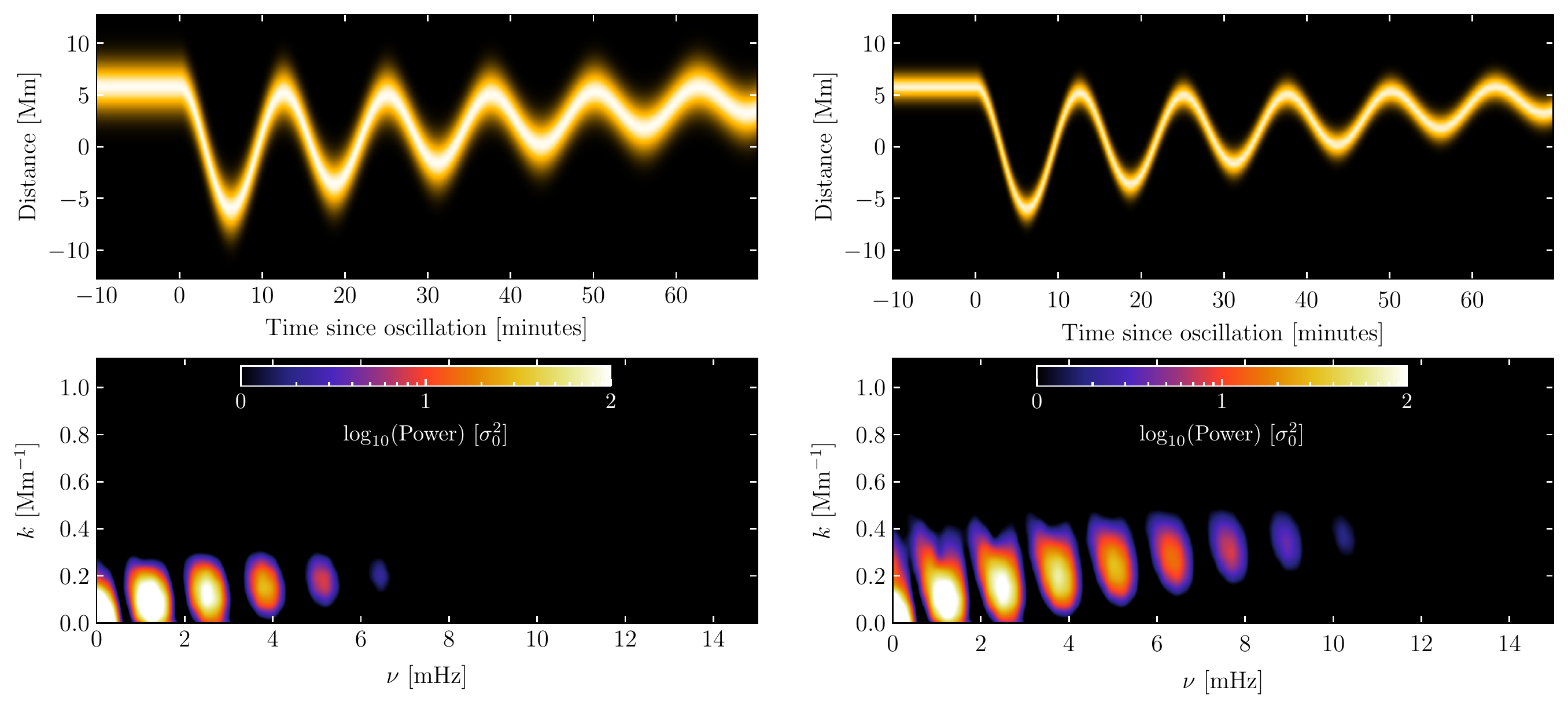}
        \caption{ {Simulated, noise-free time-distance maps (top) and their 2D power spectrum (bottom). Left panel: A thick (FWHM = 3.0 Mm) coronal loop oscillating at a single frequency of 1.33 mHz. Right panel: Equivalent to the left panel but for a thin (FWHM = 1.5 Mm) loop. The relative increase in power at high frequencies, compared to the thick loop, arises due to the Fourier transform of a narrow Gaussian. In both cases, the harmonics arise owing to the periodic but non-sinusoidal waveform of the loop.}    }   
        \label{FigureSynthetic} 
\end{figure*}

The purpose of this appendix is to demonstrate an ambiguity of identifying frequencies, and in particular harmonics, within observations of coronal loop oscillations. We show the presence of these frequencies, and the overall distribution of the power spectrum, is indirectly influenced by the non-uniform brightness of the loop. 

Figure \ref{FigureSynthetic} shows the time-distance maps of synthetic loops (top) and their 2D power spectrum (bottom). The loop in the left panel has a full width at half maximum (FWHM) of 3.0 Mm and the right panel has a width of 1.5 Mm. Both loops oscillate sinusoidally with a single frequency of 1.33 mHz. It can be determined that the spectrum of a bright loop oscillating with one frequency also contains power at several integer multiples of the fundamental mode.  However, these harmonics are artificial and can be explained as follows: Since the synthetic loop is modelled to have a width that is Gaussian at an instant in time, then its Fourier transform yields another Gaussian in the frequency domain. A time series of the perturbed loop then contains periodic, Gaussian-like oscillations.  Indeed, as the Gaussian widths are related in the spatial and frequency domains as ${\sigma_x} {\sigma_k} \propto 1$ \cite[e.g.][]{Hartmann2007}, then the spectrum of a thin loop shows more pronounced peaks at higher harmonics (shown in the bottom panel Figure \ref{FigureSynthetic}). The relative power of each harmonic contains information about the overall spatial width of the loop. Hence, as a result of the non-uniform brightness of an oscillating loop (modelled as a Gaussian here), its frequency spectrum contains power at every integer multiple of the fundamental mode, which decreases like a Gaussian. More succinctly, power at integer multiples of the fundamental mode arises because of the periodic but non-sinusoidal waveform of the intensity time series of a loop, the shape of which is defined by its width.

\section{Signatures of high-frequency oscillations }\label{A2}

Figure \ref{Harmonic} shows a faint loop, superposed onto the bright loop, oscillating with a frequency $4$ mHz, that is at the third harmonic of the bright loop. The intensity of the background loop is $50\%$ of the foreground loop. Signatures of high-frequency oscillations are clearer in the 2D FFT spectrum, but barely within the 1D spectrum.

\begin{figure}
        \centering
        \includegraphics[width=8cm]{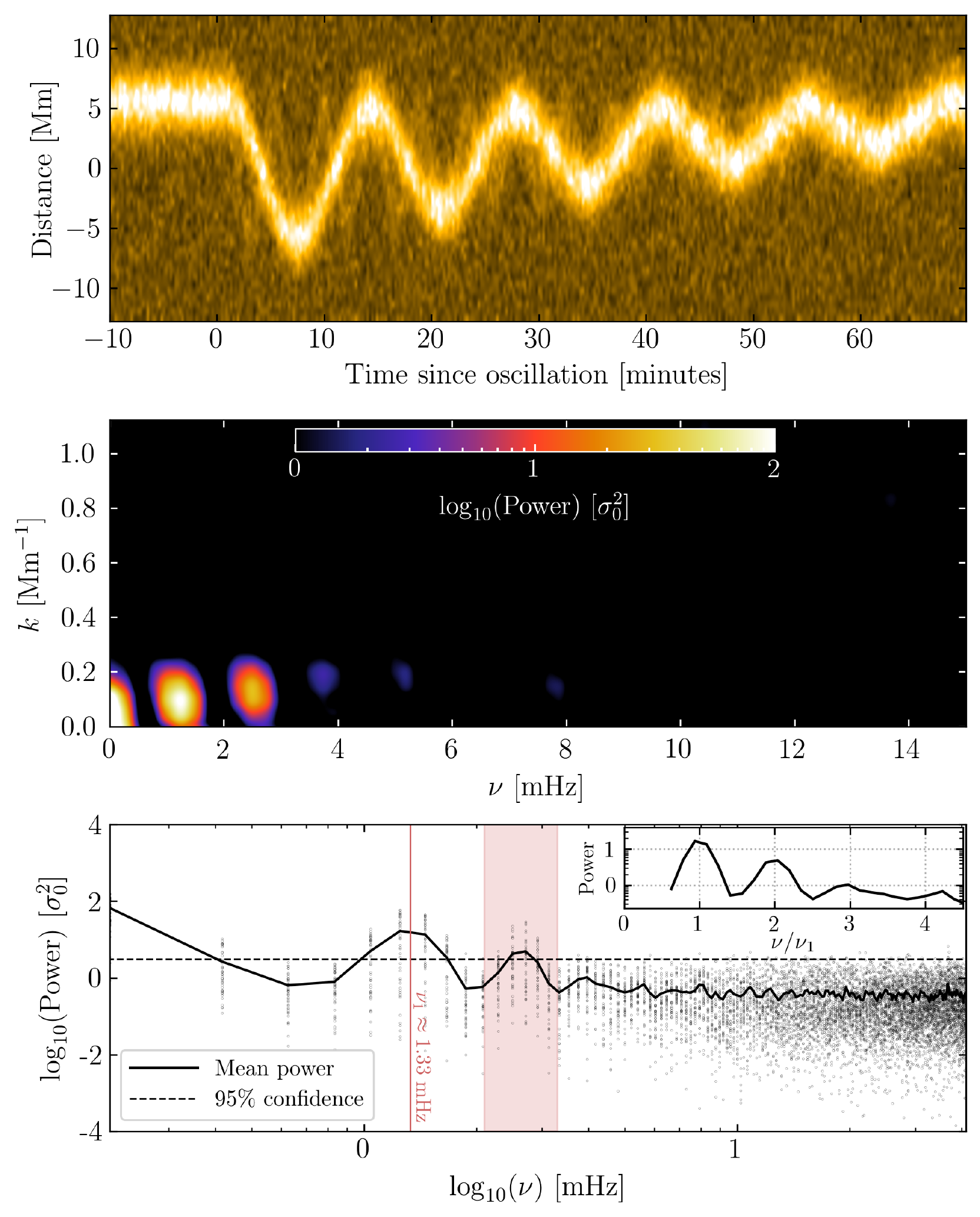}
        \caption{Same as Figure \ref{Figure4} but for a faint background loop superposed onto the bright loop. The fundamental frequency of the faint loop ($4$ mHz) is 3 times that of the bright loop. The 1D FFT spectrum is almost identical to that of the single loop in Figure \ref{Figure4}. However, the 2D spectrum shows deviations at different $k$ from the expected single loop, suggesting a high-frequency signal may be present.}   
        \label{Harmonic} 
\end{figure}

\end{appendix}

\end{document}